# Star formation in dense clusters

**Philip C. Myers**

Harvard-Smithsonian Center for Astrophysics, 60 Garden Street, Cambridge

MA 02138 USA

pmyers@cfa.harvard.edu

**Abstract.** A model of core-clump accretion with equally likely stopping describes star formation in the dense parts of clusters, where models of isolated collapsing cores may not apply. Each core accretes at a constant rate onto its protostar, while the surrounding clump gas accretes as a power of protostar mass. Short accretion flows resemble Shu accretion, and make low-mass stars. Long flows resemble reduced Bondi accretion and make massive stars. Accretion stops due to environmental processes of dynamical ejection, gravitational competition, and gas dispersal by stellar feedback, independent of initial core structure. The model matches the field star IMF from 0.01 to more than 10 solar masses. The core accretion rate and the mean accretion duration set the peak of the IMF, independent of the local Jeans mass. Massive protostars require the longest accretion durations, up to 0.5 Myr. The maximum protostar luminosity in a cluster indicates the mass and age of its oldest protostar. The distribution of protostar luminosities matches those in active star-forming regions if protostars have a constant birthrate but not if their births are coeval. For constant birthrate, the ratio of YSOs to protostars indicates the star-forming age of a cluster, typically ~1 Myr. The protostar accretion luminosity is typically less than its steady spherical value by a factor of ~2, consistent with models of episodic disk accretion.

## 1. Introduction

Understanding the origin of protostar masses is necessary for a quantitative understanding of star formation. This problem is important for star formation in clusters, where observations constrain models to produce ~ 100 protostars in a region of ~ 1 pc in a time span of ~ 1 Myr. Further, the masses of these protostars must follow the initial mass function (IMF) to within uncertainties due to statistics and protostar evolution (McKee & Offner 2010). Recent reviews of the IMF, and of star formation in clusters, are given by Bastian et al. (2010), Clarke (2010), and Lada (2010).

It is generally accepted that "isolated" low-mass protostars are born in "dense cores," or condensations which have density exceeding ~ $10^4$ cm$^{-3}$ and which appear nearly isothermal and self-gravitating (Myers & Benson 1983, Beichman et al. 1986, Shu et al. 1987). Isolated protostars are often modelled as accreting due to the gravitational collapse of such cores, starting with the pioneering studies of Larson (1969) and Shu (1977). Such isolated models are useful provided the cores are well-separated from their neighbors and provided the core self-gravity and thermal pressure are dominant over all other forces, which are then safely neglected.

This paradigm of isolated star formation applies to low-mass stars which are single or in sparse groups. Models of this process are well-developed for many initial configurations, and take into account the effects of turbulence, rotation, magnetic fields, and complex geometry (McKee & Ostriker 2007).

In some isolated models, the initial condensation has a well-defined mass boundary, and the mass of the resulting protostar is assumed to be a fixed fraction of the initial condensation mass. This picture is supported by the similarity between the distribution of core masses and the IMF (Motte et al. 1998; Alves et al. 2007). The ratio of final protostar mass to initial core mass, called the "star formation efficiency," is significantly less than unity, and its physical basis is generally ascribed to gas dispersal by outflows (e.g. Matzner & McKee 2000). Such fixed-mass models are sometimes called "monolithic collapse" (MC) models, or "turbulent core" models (McKee & Tan 2003).

Isolated models appear useful to describe star formation in the less crowded parts of clusters. In the parsec-scale filaments which extend from cluster centers, protostars can be sufficiently far from their neighbors, and dense core self-gravity can dominate over other forces due to the surrounding complex. Isolated models of initial condensations in such cluster regions are presented in Myers (2011, hereafter Paper 1).

However in the densest parts of clusters, the closer proximity of protostars, the more complex structure of their dense gas, and their turbulent motions challenge the idea that each protostar gains its mass solely from an initial condensation in a static medium of lower density.

Instead, some stars may form in the dense parts of clusters primarily by gravitational accretion from their surrounding medium, and by competition with nearby accretors (Bonnell et al. 1997). In this picture of "competitive accretion" (CA), the final mass of a protostar is set more by its history of accretion from an extended clump which harbors many cores and protostars, than by the available mass of an isolated core. The CA picture has been developed primarily by numerical simulations of cluster-forming regions. The MC and CA models are compared in more detail by Peters (2010), by McKee & Offner (2010) and by Offner & McKee (2011, hereafter OM11).

These results suggest that a simple model is needed for the dense parts of clusters, where protostars start accreting in condensations resembling dense cores, where they can also gain mass from the core environment, where their accretion durations are specified, and where protostar mass is not tied to the gravitational collapse of an isolated initial condensation. This paper gives such a description of the masses of protostars in clusters, in terms of their mass accretion rate and duration. This description is then shown to fit observed mass and luminosity distributions, and is used to estimate cluster ages.

The paper has five sections. Section 2 describes the models of mass accretion rate, and of equally likely stopping, and predicts distributions of protostar mass and accretion luminosity. Section 3 presents parameters which give the best match of the mass function to the IMFs of Kroupa (2002) and Chabrier (2005), and the best match of the accretion luminosity function to the distribution of protostar luminosities in nearby clouds (Dunham et al. 2010). Section 4 describes limitations and implications of the model, and discusses implications of assuming that the distribution of accretion durations arises instead from the infall of isolated cores. Section 5 summarizes the conclusions.

**2. Mass accretion rate model**

The best-known model of star-forming accretion is the gravitational collapse of a singular isothermal sphere (SIS, Chandrasekhar 1939). Its mass accretion rate is $\dot{m} = 0.975\sigma^3/G$, independent of protostar mass, where $\sigma$ is the isothermal sound speed and G is the gravitational constant (Shu 1977).

When the accreting medium has complex 3D spatial structure, it is more useful to model the 1D mass accretion rate with time rather than the gravitational collapse of a particular initial spatial structure. Such models of protostar accretion rate have been advanced in several forms.

Zinnecker (1982) assumed accretion of a point mass from an infinite uniform medium with negligible self-gravity, as in Bondi (1952). The protostar mass accretion rate $\dot{m}$ is then proportional to $m^p$ where m is the protostar mass and where $p = 2$. In the limit of large $m$, the mass function $mdN/dm$ for this process follows the power law $m^{-\Gamma}$ with $\Gamma = 1$, similar to the power-law dependence $\Gamma = 1.35$ found for the initial mass function by Salpeter (1955).

However, this pioneering model requires specification of an initial distribution of protostar "seed" masses, it does not indicate how accretion stops, and it does not take into account the association of protostars with dense cores in star-forming regions.

Bonnell et al. (2001) proposed that the mass accretion rate depends on the tidal radius for low-mass stars in a gas-rich cluster, and on the Bondi radius for more massive stars in cluster centers, so that $\Gamma = 1/2$ for low-mass stars and $\Gamma = 3/2$ for more massive stars. Bate & Bonnell (2005) proposed that seed masses accrete at a fixed rate from the clump environment, and are then ejected from their dense gas supply by dynamical interactions.

A twofold model of turbulent accretion for massive star formation was proposed by Krumholz et al. (2006), where Bondi accretion is assumed, with the Bondi velocity dispersion set either to the turbulent velocity dispersion or to the velocity dispersion due to vorticity.

Models of accretion onto a star-forming core have been proposed to follow $p = 1$ (Myers 2000, Basu & Jones 2004). As with earlier models, it was necessary to specify a relatively narrow initial distribution of core masses, which then broadens due to accretion. In combination with equally likely stopping, these give a power-law tail whose slope is equal to the ratio of the time scales for growth by accretion and stopping of accretion.

In the semi-analytic model of Dib et al. (2010), a clump with density varying as radius to the -2 power harbors cores whose density varies as radius to the -4 power. Cores have a constant birthrate, with a mass distribution due to turbulent fragmentation. A core grows by accretion due to clump turbulence, until it meets a timescale criterion. Then a protostar appears, with mass equal to 0.1 of the core mass. Massive star winds gradually disperse the clump gas.

The present model differs from the foregoing accretion models. Its physical setting is a cluster-forming clump which harbors multiple cores. Each protostar starts to form in a core, with a constant accretion rate. Surrounding clump gas also accretes onto the protostar. Accretion from the clump increases with protostar mass, as in reduced Bondi accretion, and eventually overtakes accretion from the core as the main source of protostar mass.

Accretion stops when the mass accretion rate becomes negligibly small, due to dynamical ejection, competition with nearby accretors, and gas dispersal by stellar feedback. These factors are considered more important than initial core structure, as discussed in Section 4.3. The termination of accretion is described statistically by assuming that accretion is equally likely to stop at any moment. Accretion is assumed to stop abruptly with no tapering, to keep the model simple and to minimize the number of parameters.

The present model is most similar to recent models in McKee & Offner (2010), OM11, Myers (2010), and Paper 1. These are compared in detail in Section 4.7, taking into account both formulation and results.

2.1. Mass and mass accretion rate

The mass accretion rate $\dot{m}$ onto a protostar of mass $m$ is written as the sum of "core-fed" and "clump-fed" components which resemble isothermal collapse and Bondi accretion,

$$\dot{m} = \dot{m}_{core} + \dot{m}_{clump} \qquad (1)$$

where $\dot{m}_{core}$ is a parameter, independent of time and protostar mass, and where $\dot{m}_{clump}$ increases as a power $p$ of protostar mass, according to

$$\dot{m}_{clump} = \dot{m}_{core}\left(\frac{m}{\dot{m}_{core}\tau_{clump}}\right)^p . \qquad (2)$$

Here $\tau_{clump}$ is a time scale for accretion of clump gas onto the protostar. This parameter is also independent of time and protostar mass. The two components of $\dot{m}$ are written in terms of free parameters to allow fits to mass and luminosity distributions, and to allow contributions to the accretion rate beyond those due to simple collapse models.

An earlier effort in this direction is the extension of the collapse of the singular isothermal sphere from rest (Shu 1977) to include the effects of nonzero initial motions, inferred from observations of starless dense cores (Fatuzzo et al. 2004).

The exponent $p$ is assumed to lie in the range 0-2, to span well-known models of accretion due to collapse of a singular isothermal sphere (Shu 1977; hereafter Shu accretion), and spherical accretion onto a point mass from a uniform isothermal medium (Bondi 1952; hereafter Bondi accretion).

In equations (1)-(2), the relative importance of the two components of mass accretion rate changes as the protostar mass $m$ increases with respect to the characteristic mass $m_0$, defined by

$$m_0 \equiv \dot{m}_{core} \tau_{clump} \quad . \tag{3}$$

When $m \ll m_0$, the core term dominates the accretion. If the core accretion onto the protostar is due to Shu accretion, then $\dot{m}_{core}$ can be expressed in terms of the initial core temperature.

Conversely, when $m \gg m_0$ the clump term dominates the accretion. Then if $p = 2$ the mass accretion rate has the form of Bondi accretion, and $\tau_{clump}$ can be expressed in terms of the temperature and density of the surrounding gas.

This two-component accretion rate can apply to many initial configurations. In the simplest spherically symmetric case, it corresponds to initial core-clump structure where the slope of the density profile changes from steep at small radii in the core, to shallow at large radii in the clump.

If $0 < p < 2$ the clump component of accretion is referred to here as "reduced Bondi accretion." Such accretion has a more gradual increase of mass with time than does Bondi accretion. It been used to model the growth of supermassive black holes (Perna et al. 2003). Modelling accretion onto supermassive black holes appears to require a compromise between Bondi accretion and free fall collapse (Hobbs et al. 2011).

Equation (1) can be put in dimensionless form by defining the dimensionless time since the start of accretion $\theta = t/\tau_{clump}$ and the dimensionless protostar mass $\mu = m/m_0$. Then

$$\frac{d\mu}{d\theta} = 1 + \mu^p . \qquad (4)$$

Integrating equation (4) relates the accretion duration $t$ to the protostar mass $m$ at $t$,

$$t = \tau_{clump} \int_0^{m/m_0} \frac{d\mu}{1 + \mu^p} . \qquad (5)$$

The integral in equation (5) has simple analytic solutions

$$m = m_0 \left[\exp(\theta) - 1\right] \qquad (6)$$

for $p=1$, and

$$m = m_0 \tan(\theta) \qquad (7)$$

for $p = 2$.

The increase of protostar mass with accretion duration is shown in Figure 1 for these two cases with parameter values $\dot{m}_{core} = 1.7 \times 10^{-6}\ M_\odot\ \text{yr}^{-1}$ and $\tau_{clump} = 0.20$ Myr. These parameter choices result from fitting the model mass distributions to the IMF, and are justified in Section 3. They imply $m_0 = 0.34\ M_\odot$ from equation (3).

Figure 1 also shows protostar mass as a function of accretion duration for the corresponding one-component core model, which is equivalent to the two-component model when $\tau_{clump}$ becomes infinite. For both core accretion and core-clump accretion, the protostar mass increases linearly with time, as $\dot{m}_{core}\ t$, for short durations when $t \ll \tau_{clump}$ or equivalently when $m \ll m_0$. In contrast, for core-clump accretion the protostar mass increases more rapidly with time for long durations when $t \gg \tau_{clump}$ or $m \gg m_0$. For such long durations the two-component models yield significantly greater protostar mass than does the one-component core model.

The increased mass accretion rate provided by the clump component in this model is needed to provide stars of sufficiently high mass, in the time periods associated with star formation in nearby star-forming regions. Models of low-mass star formation, such as Shu accretion, cannot by themselves provide enough mass in the few 0.1 Myr available. Initial equilibrium condensations described by a polytropic index less than unity can provide a greater mass accretion rate when they collapse, than do isothermal condensations (Fatuzzo et al. 2004). Models of massive star formation, based on collapse of nonisothermal condensations, have been discussed by many authors (e.g. Adams & Fatuzzo 1996, McLaughlin & Pudritz 1996, McKee & Tan 2003).

The steeper rate of increase of mass with accretion duration for high masses than for low masses is a characteristic feature of the two-component model. This property is present in models of collapsing two-component condensations such as the thermal-nonthermal "TNT" model (Myers & Fuller 1992) and the two-component turbulent core "2CTC" model (McKee & Tan 2003).

An extreme form of this steepening occurs when $p = 2$. Then Figure 1 and equation (7) indicate that the increase of mass with duration becomes infinitely steep, and all massive stars have the same maximum accretion duration $t_{max} = (\pi/2)\tau_{clump}$. For the value of $\tau_{clump} = 0.20$ Myr assumed here, this maximum duration is $t_{max} = 0.31$ Myr. This behavior is called "constant time" accretion by McKee & Offner (2010).

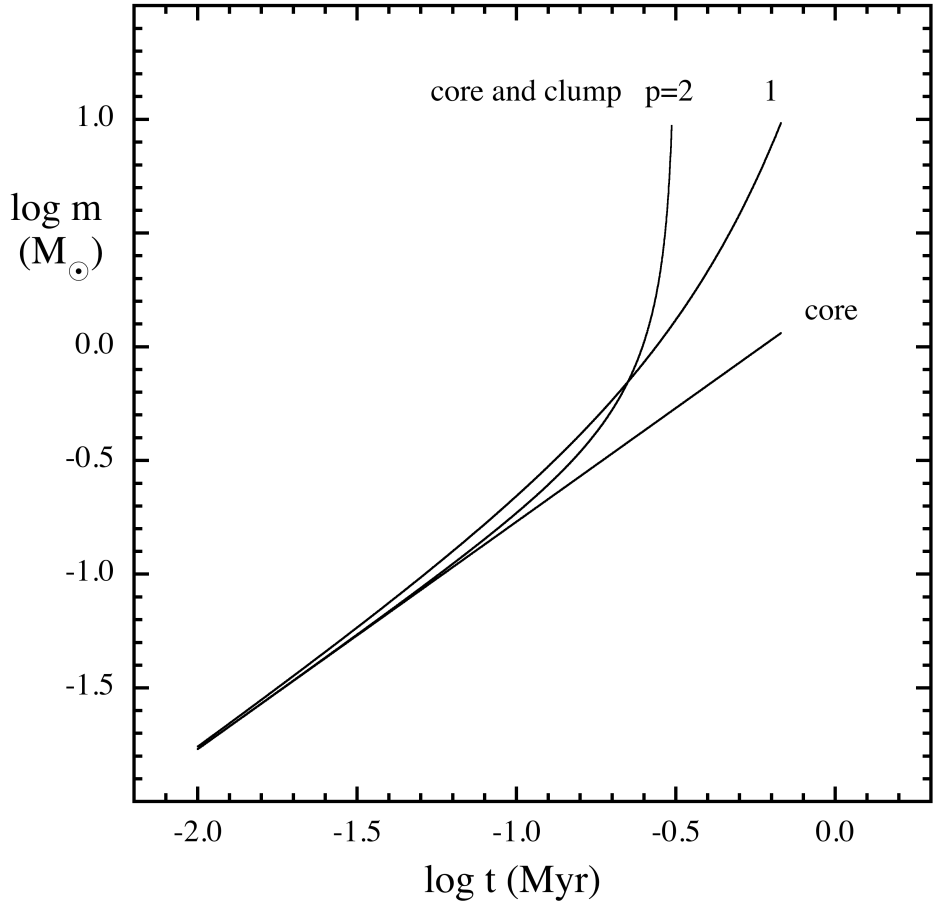

**Figure 1.** Protostar mass as a function of time for two-component models of the mass accretion rate *(core and clump)*, where the clump component depends on protostar mass as $m^p$, with $p = 1$ and $p = 2$, and where the clump component is absent *(core)*. The parameters are $m_0 = 0.34\, M_\odot$ and $\tau_{clump} = 0.20$ Myr.

The two-component model presented here is also distinctly different from the one-component clump model, where $\dot{m}_{core} = 0$. There, the protostar mass diverges if its accretion starts from the zero of time, due to the power-law form of the clump mass accretion rate. Therefore the one-component clump accretion model requires a non-zero initial "seed mass." Such seed masses were assumed to match the opacity limit on fragmentation in the clump accretion model of Bate & Bonnell (2005).

In contrast, the two-component model requires no such seed mass. It has zero protostar mass at the zero of time, because its initial accretion is core-fed rather than clump-fed, and because for core-fed accretion the protostar mass increases linearly with time.

The present model can match the accretion history of protostars in highly structured, evolving clumps, even when the spatial structure is too complex to describe analytically.

For example, the mass of the most massive protostar in the outflow-regulated, clump-fed simulation of cluster formation by Wang et al. (2010, figure 4) increases linearly with time for early times, and then more rapidly with time. In this numerical calculation, the global infall of a turbulent clump is modified by the retarding effects of magnetic forces and winds from recently formed protostars. The dense gas structure is highly filamentary and rapidly changing due to the initial clump turbulence and to subsequent shocks. The clump forms low-mass and massive protostars. For the massive protostars, much of the accretion flow is along extended filamentary paths. It seems doubtful that these complexities could be properly described by simple analytic models of 3D cluster structure and evolution. Yet the mass of the most massive protostar as a function of time is very well fit by the two-component accretion model in equation (6) for parameter values $\dot{m}_{core} = 10^{-5}\ M_\odot\ \mathrm{yr}^{-1}$ and $\tau_{clump} = 0.36$ Myr, over a protostar mass range exceeding 0-10 $M_\odot$.

2.2. Accretion luminosity

The luminosity due to accretion onto the protostar can be written

$$L_{acc} = \gamma \frac{Gm\dot{m}}{R} \qquad (8)$$

where as in Paper 1 $\gamma$ is the accretion luminosity efficiency, or the ratio of actual accretion luminosity to that for perfectly spherical steady accretion, with no absorption by the protostar of accreting internal energy. In equation (8) $R$ is the spherical radius of the accreting surface. Here the value $R=2.5R_\odot$ is adopted, following Stahler et al. (1980), Hosokawa & Omukai (2009), and OM11.

For the two-component model in Section 2.1, the accretion luminosity in equation (8) can be written

$$L_{acc} = L_0 \mu \left(1 + \mu^p\right) \qquad (9)$$

where the luminosity scale $L_0$ is defined by

$$L_0 \equiv \frac{\gamma G m_0 \dot{m}_{core}}{R} \qquad (10)$$

and the normalized mass $\mu$ is given by

$$\mu \equiv \frac{m}{m_0} \quad . \qquad (11)$$

For the parameters $m_0 = 0.34\, M_\odot$ and $\tau_{clump} = 0.20$ Myr used to calculate the variation of protostar mass with time in Figure 1, $L_0 = 7.23\gamma\, L_\odot$.

Equation (9) shows that the accretion luminosity increases with protostar mass linearly for low mass, and then more rapidly, as $m^{p+1}$ for high mass, as illustrated in Figure 2. There the value of the luminosity accretion efficiency $\gamma = 0.5$ is used, based on the model fitting described further in Section 3.

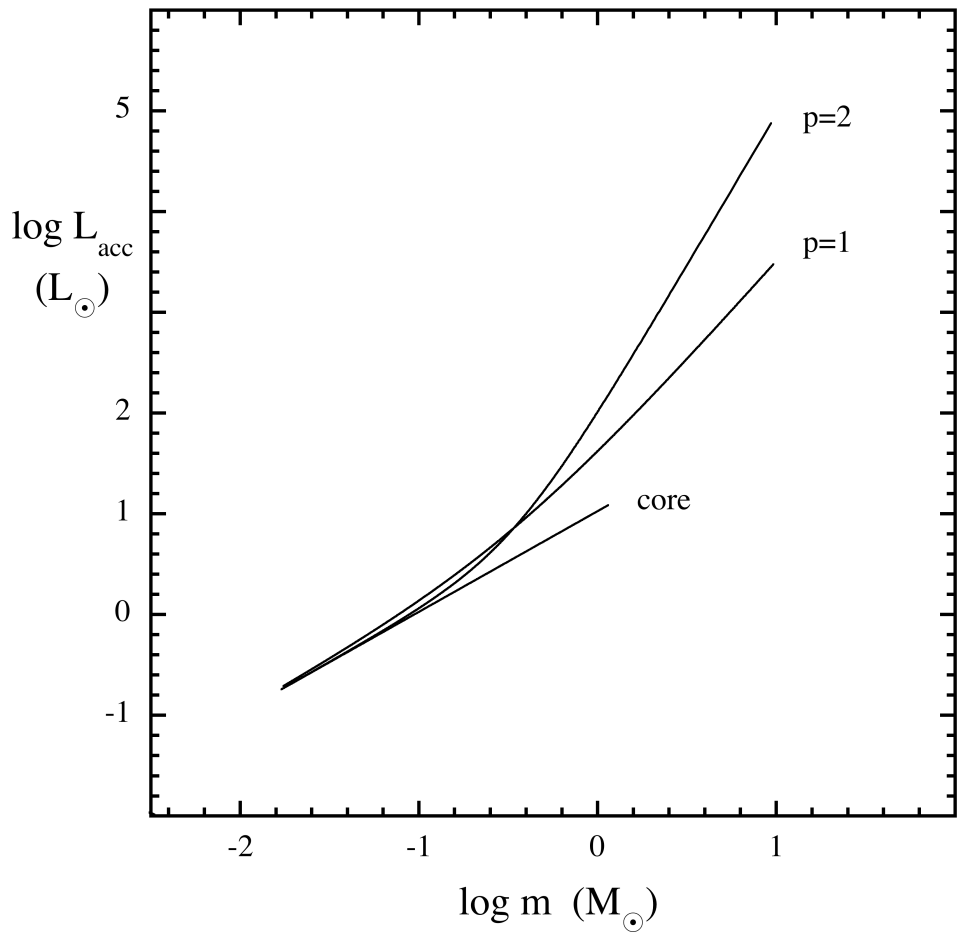

**Figure 2.** Accretion luminosity as a function of protostar mass, for the two-component accretion model of the mass accretion rate, where the clump component depends on protostar mass as $m^p$ ($p = 1$, $p = 2$), and for a one-component model of core accretion

alone *(core)*. Each curve is shown for the same range of durations as in Figure 1. The parameter values are $m_0 = 0.34\ M_\odot$ and $\tau_{clump} = 0.2$ Myr as in Figure 1, and $\gamma = 0.5$.

For protostar masses $\sim 0.2\ M_\odot$, near the peak of the IMF, Figure 2 shows that the accretion luminosity for the adopted parameters is a few $L_\odot$. This luminosity is due largely to the core component of the accretion rate. At such short accretion durations, of order 0.1 Myr, the clump contributions are smaller than the core contribution and are negligibly different between $p = 1$ and $p = 2$. In contrast, for durations approaching 1 Myr, the luminosity approaches $10^4\ L_\odot$, and the luminosity for $p = 2$ exceeds that for $p = 1$ by a factor $\sim 10$, while the luminosity from the core component alone is only $18\ L_\odot$.

The accretion luminosities in Figure 2 are much greater than the luminosities of hydrogen-burning main-sequence stars of the same mass, for the mass range 0.01 - 10 $M_\odot$ considered here. For the approximate relation between main-sequence stellar luminosity and mass $L/L_\odot = (M/M_\odot)^{3.5}$, the accretion luminosity for $p = 1$ exceeds the main-sequence luminosity by a factor of 40 for stars of solar mass, and the two luminosities become equal when $M = 10\ M_\odot$.

The accretion luminosity increases with protostar mass only while the protostar is accreting. Once the protostar has reached its final mass, the accretion luminosity decreases to zero. Thus Figure 2 should be understood as describing the accretion luminosity as a function of accreting protostar mass, or alternatively as describing the maximum accretion luminosity as a function of final protostar mass.

2.3. Equally likely stopping

To understand the distributions of protostar masses and luminosities, it is necessary to understand the distribution of accretion durations, and the physical mechanisms which limit accretion. It has been suggested that the shortest durations are due to dynamical ejections from small multiple systems (Reipurth & Clarke 2001, Bate & Bonnell 2005, Pudritz 2010). Longer durations may be due to gravitational competition for mass by nearby accretors ("competitive accretion," Bonnell et al. 1997). Longer durations may also be set by stellar feedback, which disperses dense gas otherwise available for accretion, due to winds, outflows, heating and ionization (Myers 2008, 2009).

The relative importance of these and other mechanisms for limiting accretion is poorly known. A statistical description is therefore useful until a better physical understanding becomes available. One such model assumes that accretion is equally likely to stop at any moment. Then the probability that the duration of accretion lies between $t$ and $t + dt$ is equal to the probability that accretion does not stop until $t$, $\exp(-t/\bar{t})$, times the probability of stopping between $t$ and $t + dt$, $dt/\bar{t}$, or

$$p(t)dt = \exp(-t/\bar{t})\left(\frac{dt}{\bar{t}}\right) \qquad (12)$$

where $\bar{t}$ is the mean duration of accretion. Here the durations $t$ extend from 0 to ∞. However, equation (12) is sufficiently accurate for the present application provided the maximum duration $t_{max}$ is significantly longer than the mean duration $\bar{t}$, as in section 3 where $t_{max}/\bar{t} > \sim 5$.

The durations due to equally likely stopping have probability density which declines exponentially with increasing duration. A similar decreasing exponential formulation was used to describe the duration of accretion in earlier models of core accretion, dynamical ejection of low-mass protostars, and of protostar accretion (Myers 2000, Basu & Jones 2004, Bate & Bonnell 2005, Myers 2009, 2010 and Paper 1).

This statistical description of accretion durations has been called "random stopping" in Paper 1 and by other authors, but this term does not mean that the stopping times are described by random sampling of a uniform distribution. Instead, the distribution of stopping times $t_{stop}$ is related to the accretion durations $t$ and to the distribution of starting times $t_{start}$ through the definition $t = t_{stop} - t_{start}$. If all the protostars in an ensemble start accreting at the same time $t_{start} = 0$, their distribution of $t_{stop}$ is essentially the probability of not stopping until $t_{stop}$, and it follows the decreasing exponential $\exp(-t_{stop}/\bar{t})$ from equation (12). If instead $t_{start}$ is distributed, the distribution of $t_{stop}$ depends on the distributions of $t_{start}$ and $t$, as discussed further in Sections 2.4 and 3.2.

2.4. Mass distributions

Comparison of predicted mass functions with observed mass functions is currently possible for main-sequence stars and for pre-main sequence stars (Bastian, Covey & Meyer 2010), but not for protostars, due to absorption of optical and near-infrared lines by the protostar envelope. Nonetheless, it is possible to predict protostar mass functions for different accretion models (McKee & Offner 2010), and in turn to predict corresponding protostar luminosity functions. These luminosity functions can be compared with observations, and can provide useful tests of accretion models (Kenyon et al. 1990; Fletcher & Stahler 1994a,b; McKee & Offner 2011; OM11).

2.4.1. Final mass function. The final protostar mass function, which is denoted here as $\Phi_{mf}$, is the mass distribution evaluated at a time when all the YSOs in an ensemble have completed their protostellar accretion. It depends on the distribution of accretion durations, but is independent of the distribution of accretion start times. Here $\Phi_{mf}$ is defined by $\Phi_{mf} = mp(m)$, where $p(m) = p(t)/\dot{m}$ is the probability that the protostar mass lies between $m$ and $m+dm$. This mass function is obtained from equations (5), (11), and (12) as

$$\Phi_{mf} \equiv \frac{q\mu}{1+\mu^p}\exp(-q\theta) \quad . \tag{13}$$

Equation (13) is derived using the same procedure as in Section 2.2 of Paper 1. Here $\Phi_{mf}$ is expressed in terms of the dimensionless mass $\mu$ and the integral $\theta$ over $\mu$, using equation (5). The parameter $q$ is the ratio of the clump accretion time scale to the mean accretion duration,

$$q \equiv \frac{\tau_{clump}}{\bar{t}} \quad . \tag{14}$$

When the protostar mass becomes much larger than $m_0$, or equivalently when $\mu \gg 1$, $\Phi_m$ approaches a power-law dependence on $\mu$, as $\mu^{-q}$ when $p = 1$, and as $\mu^{-1}$ when $p = 2$. In the cases considered here, a small departure from power-law shape is evident at high mass, because $\mu$ is not sufficiently large to reach these asymptotic limits.

The mass function in equation (13) has its peak value when the protostar mass $m$ is comparable to $m_0 = \dot{m}_{core}\tau_{clump}$, or equivalently to $\dot{m}_{core}\bar{t}q$. This modal mass is exactly equal to $m_0$ for $p = 1$, and decreases to about $0.5\,m_0$ as $p$ increases to 2, depending on the value of $q$. Thus the modal mass is essentially the product of the constant component of the mass accretion rate, times the mean accretion duration.

The mass function gets its shape from the dependence of protostar mass on accretion duration, and from the distribution of accretion durations. The dependence of protostar mass on duration is assumed to be negligibly different from one protostar to the next, and the probability distribution of durations is the same over the ensemble of protostars. In a more realistic treatment, the accretion rate and the likelihood of accretion stopping should depend on location in the cluster potential and on the time since the start of star formation. Furthermore the variation from core to core in temperature, turbulence, rotation, and magnetic energy should be described more realistically. The present model provides a simple point of comparison for more realistic simulations.

2.4.2. Present-day mass functions. Comparison of the model mass function with present-day protostar mass functions depends on accretion start times as discussed in Section 2.3. This subsection describes present-day mass functions for coeval star formation and for star formation distributed in time.

The simplest model of accretion start times is "coeval" star formation, where nearly all stars start accreting at the same time. Several studies of young clusters have applied models of pre-main sequence evolution to conclude that star formation is coeval within uncertainty of order 1 Myr (NGC 2362, Moitinho et al. 2001; Upper Scorpius, Slesnick et al. 2008).

For simplicity a limiting version of coeval star formation is considered here, where all stars start accreting at exactly the same instant of time.

If all stars in a cluster have coeval accretion starts and equally likely stopping, their accretion stop times follow their distribution of accretion durations, as in equation (13), up until the time at which the distribution is evaluated. Thus the protostar mass distribution at time $t_s$ since the onset of star formation is given by equation (13), for masses up to a maximum corresponding to duration $t = t_s$, plus a delta function of finite amplitude at the maximum mass, representing the protostars which are still accreting. In the present model where mass increases monotonically with duration, most low-mass stars would reach their final masses while more massive stars are still accreting.

An alternate picture of birth history is that protostar births are smoothly distributed in time, either at a uniform birthrate, or accelerating with time (Palla & Stahler 1999, 2000, Tan et al. 2006, Huff & Stahler 2006, Jeffries 2007, Reggiani et al. 2011). For such smoothly distributed births, the number of protostars born between $t'_s$ and $t'_s + dt'_s$ and still accreting at $t_s$ can be written

$$dN_{ps} = b(t'_s)dt'_s \exp\left[-(t_s - t'_s)/\bar{t}\right] \quad (15)$$

Here $b(t'_s)$ is the protostar birthrate at the earlier time $t'_s = t_s - t$, or equivalently at $t'_s = t_s - \theta\tau_{clump}$. The interval around protostar mass $m$ at $t_s$ due to the interval in birth times at $t'_s$ is

$$dm = \dot{m}_{core}dt'_s\left(1 + \mu^p\right) \quad (16)$$

based on equation (4).

If the birthrate $b(t'_s)$ has the constant value $b$, integration of equation (15) indicates that the number of protostars $N_{ps}$ approaches a constant value after a few stopping times, due to the balance between births and stopping,

$$N_{ps} = b\bar{t}\left[1 - \exp(-t_s/\bar{t})\right] \quad . \tag{17}$$

In equation (17), the number of protostars $N_{ps}$ at time $t_s$ can also be related to the maximum possible protostar mass $m_{max}$ at $t_s$. This maximum mass arises for a protostar which started accreting at the earliest possible time, $t'_s = 0$. Then equation (5) can be written

$$\theta_{max} = \int_0^{\mu_{max}} \frac{d\mu}{1 + \mu^p} \tag{18}$$

where $\theta_{max} = t_s/\tau_{clump}$ and $\mu_{max} = m_{max}/m_0$. Combining equations (15-18), the protostar mass distribution at time $t_s$ since the onset of star formation is $\Phi_m(t_s) = m(dN_{ps}/dm)/N_{ps}$, given by

$$\Phi_m(t_s) = \frac{q\mu}{1 + \mu^p} \frac{\exp(-q\theta)}{1 - \exp(-q\theta_{max})} \tag{19}$$

for masses $0 < m < m_{max}$. Note that in equation (19) $\theta_{max}$ can be written in terms of maximum mass, using equation (18), or in terms of cluster star-forming age $t_s$, since $q\theta_{max} = t_s/\bar{t}$.

Equation (19) describes the protostar mass distribution for core-clump accretion, equally likely stopping, and constant protostar birthrate. It depends on time because the maximum mass $m_{max}$ increases with the time $t_s$ since the onset of star formation. As $t_s$ becomes large compared to the mean stopping time $\bar{t}$, or equivalently as $m_{max}$ becomes large compared to $m_0$, the formation of protostars reaches a steady state and the term $\exp(-q\theta_{max})$ becomes negligible. Then the mass distribution $\Phi_m(t_s)$ approaches the time-independent final mass distribution, $\Phi_{mf}$ in equation (13).

The distribution in equation (19) has similar shape to some of the protostar mass functions presented in McKee & Offner (2010, hereafter MO10), which were obtained in a steady state model for various accretion laws. This distribution differs from those of MO10 however, in that it depends explicitly on the time since the onset of star formation. It therefore applies to young clusters which have not yet reached steady state, as well as to those in steady state. It also applies to clusters whose maximum mass extends to 10 $M_\odot$, where MO10 consider a maximum mass of 3 $M_\odot$.

For constant birthrate, the number of protostars $N_{ps}$ in equation (17) can be related to the star-forming age of the cluster. Writing the total number of stars at time $t_s$ as $N_s = b\, t_s$, and using equation (17) gives

$$\frac{N_{ps}}{N_s} = \frac{1 - \exp(-t_s/\bar{t})}{t_s/\bar{t}} \quad . \tag{20}$$

Thus the protostar fraction $N_{ps}/N_s$ decreases with increasing star-forming age $t_s$. This behavior was also seen in the model of Fletcher & Stahler (1994a, b). The protostar fraction decreases from its initial value of unity and approaches $\bar{t}/t_s$ when $t_s >> \bar{t}$. Thus a protostar fraction of 0.1 implies a star-forming age of $10\bar{t}$. For $\bar{t} = 0.1$ Myr as adopted in Section 3, the star-forming age is $t_s = 1$ Myr.

Equation (20) can be applied to the results of a recent study of nine young star-forming regions, based on *Spitzer Space Telescope* observations. There, the median protostar fraction $N_{ps}/N_s$ is 0.075 in the Orion region (Kryukova et al. 2011). If $\bar{t} = 0.1$ Myr as assumed above, the typical star-forming age in these regions is 1.3 Myr. This estimate is comparable within a factor 2 to other estimates of star-forming age in nearby complexes (Evans et al. 2009).

If the birthrate is accelerating, the present-day mass distribution differs in shape from the final mass distribution. If the birthrate is an increasing exponential as suggested by Palla & Stahler (2000), the log-log slope of the high-mass tail of $\Phi_m(t_s)$ becomes steeper than that of $\Phi_{mf}$ by the factor $1 + \bar{t}/\tau_b$, where $\tau_b$ is the exponential time scale of the birthrate. This change in shape may be relatively small. For mean stopping times of order 0.1 Myr as found in this paper, and for birthrate time scales of order 1 Myr as found by Palla & Stahler (2010), this change in slope is smaller than the usual uncertainty in observed mass function slope.

2.5. Luminosity distributions

The present-day luminosity function $\Phi_L(t_s)$ for protostars with core-clump accretion and equally likely stopping is obtained in essentially the same way as the present-day mass function in Section 2.4, with similar results.

If the protostar births are coeval, the present-day distribution of accretion luminosity is a single-valued function. After the onset of star formation, some of the young stellar objects (YSOs) have stopped accreting and have become pre-main sequence stars. Their luminosity due to pre-main sequence contraction is significantly less than their earlier luminosity when they were still accreting. The remaining protostars are still accreting. They all have the same accretion luminosity, since they all started accreting at the same time, with the same dependence of accretion rate on time. Consequently their protostar luminosity function is a finite-amplitude delta function of luminosity, whose amplitude decreases with time as its luminosity increases with time. The combined luminosity function of pre-main sequence stars and protostars may be bimodal, with a high-amplitude, low-luminosity peak for the pre-main sequence stars and a low-amplitude, high-luminosity peak for the protostars.

Recent studies of protostar luminosities find a broad distribution over a decade in luminosity (Dunham et al 2010, Kryukova et al 2011), much broader than the narrow function expected from the present model with coeval births.

If the protostar births are smoothly distributed in time rather than coeval, the present-day distribution of their accretion luminosity at time $t_s$ since the onset of star formation, $\Phi_L(t_s) = L_{acc}(dN_{ps}/dL_{acc})/N_{ps}$, is obtained as for $\Phi_m(t_s)$ in equation (19). Combining the number of protostars $dN_{ps}$ born between $t'_s$ and $t'_s + dt'_s$ which are still accreting at $t_s$ with the accretion luminosity interval between $L_{acc}$ and $L_{acc} + dL_{acc}$ for a protostar at $t_s$ gives

$$\Phi_L(t_s) = \frac{q\mu}{1+(1+p)\mu^p} \frac{\exp(-q\theta)}{1-\exp(-q\theta_{max})} \quad . \quad (21)$$

Here as for the mass distribution in equation (19), the accretion luminosities at $t_s$ extend from 0 to a maximum possible value $L_{max}$. This maximum value corresponds to a protostar which began accreting at the onset of star formation, $t'_s = 0$, and is still accreting at $t_s$. Its luminosity is

$$L_{max} = L_0 \mu_{max}(1 + \mu_{max}^p) \qquad (22)$$

based on equations (9) and (18). For the model parameters adopted in Section 3, a cluster whose first protostar began accreting 0.5 Myr ago has a maximum possible present-day mass 6.8 $M_\odot$ and maximum possible luminosity 2700 $L_\odot$ provided it is still accreting.

As with the mass distribution, the time-dependent present-day distribution of accretion luminosities $\Phi_L(t_s)$ in equation (21) approaches a time-independent form $\Phi_L$ when $t_s \gg \bar{t}$ or equivalently when $L_{max} \gg L_0$,

$$\Phi_L = \frac{q\mu}{1 + (1+p)\mu^p} \exp(-q\theta) \quad . \qquad (23)$$

This luminosity function depends on one variable, the accretion luminosity $L_{acc}$, and on the parameters $p$, $q$, $m_0$, and $L_0$. For simplicity, in equation (23) $L_{acc}$ is written in terms of the normalized mass $\mu$, using equation (9), to show the similarity between the luminosity and mass functions. Each distribution function increases linearly with $\mu$ for $\mu \ll 1$, rising toward a local maximum when $\mu$ is of order unity, and then declines for $\mu \gg 1$, with shallower slope than for $\mu \ll 1$. These shape properties are similar to those of present-day distributions of accretion luminosity calculated for three accretion models, whose distributions of final protostar mass match the IMF (OM11).

## 3. Comparing model distributions with observations

### 3.1. Mass function

The protostar final mass function in equation (13) was compared with formulations of the field star IMF due to Kroupa (2002) and Chabrier (2005) by varying the parameters $p$, $m_0$, and $q$ for the best fit by eye, over the mass range 0.01 to 10 $M_\odot$. In general, acceptable fits were obtained over the range of exponents $p = 1 - 2$. Those with $p = 1$ give the best match to the high-mass tail with log-log slope $\Gamma = 1.35$ (Salpeter 1955; Bastian et al. 2010), but they tend to be too broad. Those with $p = 2$ have too much curvature at high mass to match the Salpeter slope.

For each trial value of $p$, the parameter $m_0$ was adjusted mainly to shift the peak of the mass function and $q$ was adjusted mainly to change the slope of the high-mass tail. The high-mass tail is not a pure power law except for $p=1$, so in general best-fit values of $q$ do not match the Salpeter slope of 1.35.

The best-fit mass functions have $p = 1.2$, $q = 2.0$, and $m_0 = 0.34\ M_\odot$, and $p=1.3$, $q= 2.2$, and $m_0 = 0.40\ M_\odot$, as shown in Figure 3. These curves have the same modes as the reference IMF curves, at 0.16 $M_\odot$, and their low-mass and high-mass tails lie between the corresponding IMF tails. In Figure 3 the segmented IMFs of Kroupa (2002) and Chabrier (2005) are replaced by continuous approximations (Myers 2010, Appendix), because it is easier to match the mass function model to the continuous approximation than to the segmented original curve.

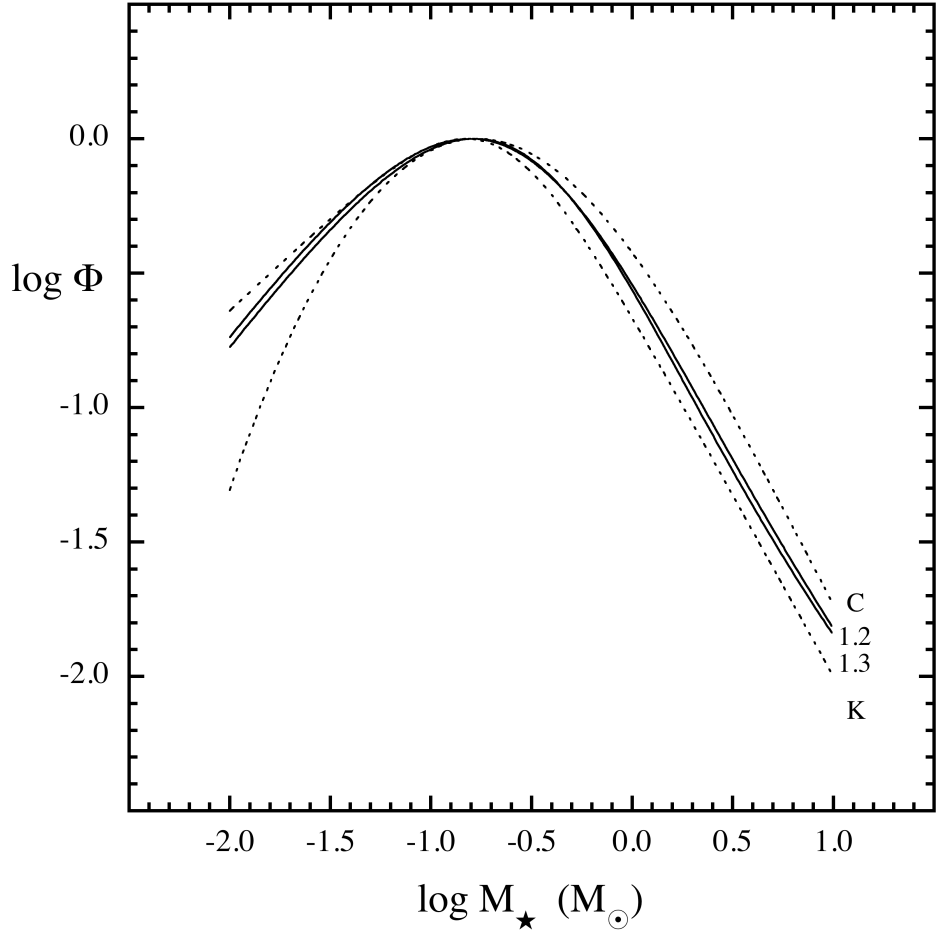

**Figure 3.** Protostar mass functions for models based on equation (15) for parameters $p$ =1.2, $q$ = 2.0, and $m_0$ = 0.34 $M_\odot$, and for $p$ =1.3, $q$ = 2.2, and $m_0$ = 0.40 $M_\odot$ *(solid lines)*, compared with estimates of the field star IMF by Kroupa (2002) and Chabrier (2005), using the continuous approximations in Myers (2010, Appendix) *(dotted lines)*.

The quality of the fits in Figure 3 is indistinguishable between $p$ = 1.2 and $p$ = 1.3, and this similarity suggests that the uncertainty in $p$ for best fit is of order 0.1. To proceed with a definite result, the value $p$ = 1.2 is adopted henceforth.

The fitting of the protostar mass function in Figure 3 sets the parameters $p$, $q$, and $m_0$. To fully specify the model it is also necessary to specify the core mass accretion rate $\dot{m}_{core}$, the time scale $\tau_{clump}$ for accretion from the clump onto the protostar, and the mean accretion duration $\bar{t}$, using equations (3) and (14). These properties are discussed in Section 3.2, once the luminosity function has also been fit.

3.2. Luminosity function

Luminosity functions of protostars can test models of protostar accretion against observations, since accretion luminosity is sensitive to the product of protostar mass and mass accretion rate (Kenyon et al 1990, Fletcher & Stahler 1994a,b, McKee & Offner 2011, OM11). Such tests are important to discriminate between competing models.

The time-independent distribution of present-day accretion luminosity in equation (23) is compared in Figure 4 to a recent compilation of 112 luminosities in nearby star-forming clouds. These luminosities were obtained by integration over the broadband spectra of 39 Class 0 and 73 Class I protostars observed primarily in the Serpens, Ophiuchus, and Perseus complexes (Dunham et al 2010). The observations were obtained with the *Spitzer Space Telescope* under the "c2d" program (Evans et al. 2009) and with the "bolocam" array at the Caltech Submillimeter Observatory (Enoch et al 2009). The distribution is notable for its breadth, about an order of magnitude in luminosity. Distributions of similar breadth are also seen in the recent study of protostar luminosities in regions forming more massive stars (Kryukova et al. 2011).

These distributions of observed protostar luminosities are much broader than expected for a model of strictly coeval births, as noted in Section 2.5. Therefore models with distributed protostar births are considered in more detail here. For simplicity a constant birth rate is assumed, since fewer parameters are needed than for a model of accelerated births. Then the width and shape of the luminosity function are completely determined by the parameter values $p$, $m_0$, and $q$ used to match the final protostar mass function to the IMF.

The fit was obtained by adjusting the accretion luminosity efficiency $\gamma$ defined in equation (10) to the value $\gamma = 0.5$. Adjusting $\gamma$ shifts the distribution as a function of luminosity but does not change its shape.

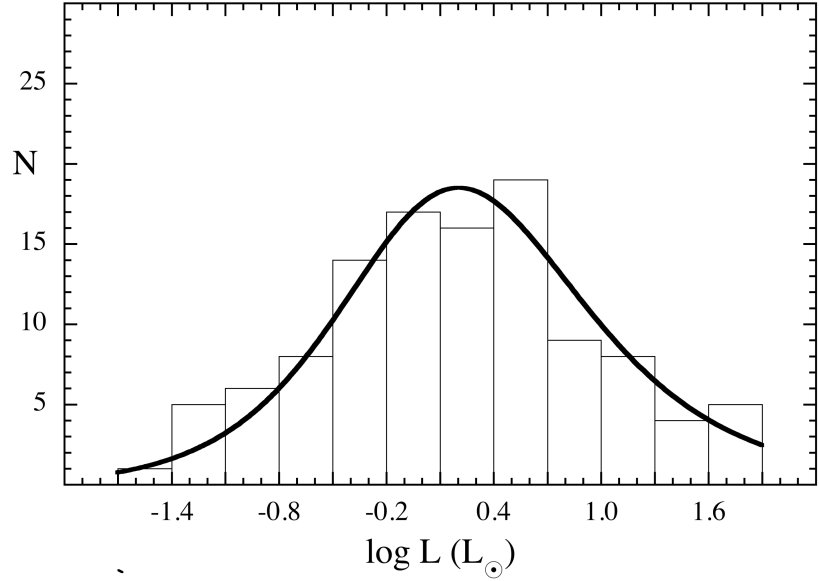

**Figure 4.** Distribution of accretion luminosity, based on equation (23) and parameters set by mass distribution in Figure 3, assuming constant protostar birthrate and $p = 1.2$, when the luminosity scale factor in equation (12) has the value $L_0 = 3.62\ L_\odot$. This distribution is superposed on the distribution of 112 values of bolometric luminosity for protostars observed in nearby star-forming regions (Dunham et al. 2010).

With this choice of $\gamma$, the luminosity scale factor defined in equation (10) has the value $L_0 = 3.62\ L_\odot$ and the modal value of accretion luminosity is $1.59\ L_\odot$. The fit is relatively good, since for each histogram bin, the model curve does not deviate by significantly more than the $\sqrt{N}$ error for Poisson statistics. However the model also predicts that this sample should have one or two protostars with luminosity in the range 80-160 $L_\odot$, which were not observed. Furthermore, this sample of protostars is limited by selection effects and small sample size, as discussed in Section 4.1.

Specification of the model parameters can now be completed, by combining the fit values of $p$, $m_0$, and $q$ from the mass function and the fit value of $L_0$ from the luminosity function, with constraints on the parameter $\tau_{clump}$ to about 0.2 Myr, due to estimates of the core gas temperature and on the clump column density.

The first constraint comes from equating the constant component of the mass accretion rate to that of Shu accretion for the collapse of a singular isothermal sphere (Shu 1977). Then the corresponding initial core temperature $T_{SIS}$ is given by

$$\frac{T_{SIS}}{K} = 10.6 \left(\frac{\tau_{clump}}{0.2 \text{ Myr}}\right)^{-2/3}. \quad (24)$$

This result suggests that if $\tau_{clump}$ gets much larger than 0.2 Myr, the effective core temperature will fall below the kinetic temperature of most dense cores, according to observations of ammonia lines (Jijina et al. 1999).

A second constraint comes from assuming that the time scale for accretion from the clump onto the protostar is comparable to the free fall time of a uniform spherical clump of radius $R_{clump}$. Then the mean column density of that clump has the value

$$\frac{\overline{N}_{clump}}{10^{22} \text{ cm}^{-2}} = 5.89 \left(\frac{R_{clump}}{0.5 \text{ pc}}\right)\left(\frac{\tau_{clump}}{0.2 \text{ Myr}}\right)^{-1/2} \quad (25)$$

This result indicates that if $\tau_{clump}$ gets much smaller than 0.2 Myr, the mean clump column density will substantially exceed that of most embedded clusters within 1 kpc of the Sun (Gutermuth et al. 2008).

Taken together, equations (24) and (25) suggest that 0.2 Myr is a reasonable choice for $\tau_{clump}$. Then equations (3), (10), and (14) imply that the constant component of the mass accretion rate is

$$\dot{m}_{core} = 1.7 \times 10^{-6} \; M_\odot \; \text{yr}^{-1}, \quad (26)$$

and the mean accretion duration is

$$\bar{t} = 0.1 \; \text{Myr} \; . \quad (27)$$

These parameters indicate that the accretion duration for a 10 $M_\odot$ protostar is 0.54 Myr.

3.3. Parameter uncertainties

Parameter uncertainties are estimated from the variation in parameter value required to give a noticeable decrease in the quality of the fit in Figure 3 from the best-fit values $p = 1.2$, $q = 2.0$, $m_0 = 0.34 \; M_\odot$, and in Figure 4 from $\gamma = 0.5$. These uncertainties are $\sigma(p) = 0.1$, $\sigma(q) = 0.2$, $\sigma(m_0) = 0.06 \; M_\odot$, and $\sigma(\gamma) = 0.05$. The uncertainty in $\tau_{clump}$ is estimated as $\sigma(\tau_{clump}) = 0.05$ Myr, which would give a significant decrease in core temperature below values observed in clusters, as discussed in Section 3.2 above.

Assuming the above uncertainties are independent, they are combined to yield uncertainties $\sigma(\dot{m}_{core}) = 0.5 \times 10^{-6}\ M_\odot\ yr^{-1}$ and $\sigma(\bar{t}) = 0.03$ Myr.

These estimated uncertainties are based primarily on fits to observed distributions, but not on the observational uncertainties in those distributions. Therefore the above uncertainty estimates should be considered lower limits, possibly by a factor of order 2. A better estimate of parameter uncertainties should be based on more detailed studies of mass and luminosity functions in young clusters.

3.4 Implications of fit parameter values

The fitting of the IMF and luminosity functions sets three of the model parameters to values characteristic of local star-forming regions. The constant component of the accretion rate $\dot{m}_{core}$ corresponds to the infall of a singular isothermal sphere whose temperature is 11 K. This temperature is similar to that of dense cores in the regions studied by Evans et al. (2009) and Dunham et al. (2010), according to observations of lines of $NH_3$ (Jijina et al. 1999). The time scale of the mass-dependent component of the accretion rate, $\tau_{clump} = 0.2$ Myr, can be interpreted as a free-fall time for clump gas. Then the mean clump density is $3 \times 10^4\ cm^{-3}$, typical of the intercore dense gas in the Oph B region (Friesen et al. 2009). The luminosity accretion efficiency is 0.5, suggestive of disk accretion with a modest degree of episodic accretion (OM11).

The fitting of the IMF and luminosity functions may also improve our understanding of accretion and of cluster ages. The best-fit value of the mass exponent in the clump component of the accretion rate, $p = 1.2$, indicates that the physical nature of the accretion differs from that of pure Bondi accretion, for which $p = 2.0$. Assuming cold spherical clump infall from rest, this accretion rate mass exponent requires the clump gas to be more centrally condensed than a uniform medium, but less centrally condensed than an isothermal sphere. This result corroborates the clump radial structure derived by assuming the IMF as an input to a model of cold spherical infall and equally likely stopping, shown in Figure 4 of Myers (2010). Together these results favor accretion by centrally condensed gas with $p = 1.2$ over uniform accretion with $p = 2.0$. Such central condensation appears to be a more realistic description of radial clump structure than the uniform medium of pure Bondi accretion.

The mean accretion duration $\bar{t}$ may be a useful indicator of the star-forming age of a cluster $t_s$, as suggested by the relation between $\bar{t}$, $t_s$, and protostar fraction in equation (20). As more accurate population studies of embedded clusters become available, it may become possible to compare cluster ages from pre-main sequence tracks with cluster ages based on protostar fraction models. Such a comparison may lead to a better understanding of protostar birth history in clusters.

Similarly, the parameters $p$, $\gamma$, $\dot{m}_{core}$, and $\tau_{clump}$ which set the accretion luminosity also set the age of the oldest protostar in a cluster, $t_{ps}$, which can be obtained from the luminosity of the most luminous protostar and equations (5) and (9-11). The age of the oldest star should be an upper limit on the age of the oldest protostar, $t_s \geq t_{ps}$. For a cluster with a total of 100 YSOs, including 10 protostars, whose most luminous protostar has $10^3 \, L_\odot$, the present model and its fit parameter values imply $t_s = 1$ Myr and $t_{ps} = 0.6$ Myr, satisfying the expected relation $t_s \geq t_{ps}$. The discussion of uncertainties in Section 3.3 implies that these ages are each uncertain by factors 1.3-1.6.

## 4. Discussion

The foregoing sections show that a core-clump model of protostar accretion flow, with equally likely stopping, can match observed distributions of protostar mass and accretion luminosity, independent of assumptions about core masses, star formation efficiency, initial seed masses, or clump gas structure.

Such freedom is useful, since observations and simulations suggest that dense cluster-forming gas may be too complex to describe with simple spatial models. Furthermore, a given accretion history may arise from more than one structure model, depending on the relative importance of magnetic fields, turbulence, rotation, and stellar feedback.

The parameter values resulting from these fits are consistent with temperatures of core gas and density of clump gas inferred from observations. The luminosity accretion efficiency is consistent with a modest degree of episodic disk accretion. The accretion rate mass exponent indicates that the clump gas has a centrally condensed radial structure. The mean accretion duration provides a new way to estimate the star-forming age of a cluster, based on the fraction of all YSOs which are still accreting. The parameters which set the accretion luminosity also provide an estimate of the age of the oldest protostar in the cluster.

This section discusses some limitations and implications of these results.

4.1. Limitations

The models presented here are consistent with two important distributions - the IMF and protostar luminosities. The best-fit parameters describing core gas temperature and star formation time also have plausible values. It remains to go beyond these points of consistency, to make predictions which could better discriminate among different models of protostar birth history.

The models described here should be formulated to predict distributions of accretion luminosity for a variety of protostar birth histories, including coeval births, uniform birth rate, accelerating births, and the deceleration in births due to dispersal of cluster gas. Then the best fit among models of birth history can be identified, rather than in the present case where one model is shown to have a good fit. In addition, comparison with other accretion models should be made, as has been done recently by OM11.

A key component of the present models is the assumption that the duration of accretion is distributed due to multiple causes, including dynamical ejection, competition from other accretors, and dense gas dispersal due to stellar feedback. This assumption is plausible, but only because our present knowledge of accretion durations is so limited. It may be useful to analyze numerical simulations of cluster formation, such as those of Wang et al. (2010) and others described by Pudritz (2010), to compare their distributions of accretion duration, and to evaluate the effect of varying distributions of accretion duration on the parameters needed to fit mass and luminosity functions.

In the present model, accretion durations have a single probability distribution, with a single parameter, the mean accretion duration, over the star-forming life of the cluster and throughout its volume. This simple formulation cannot account for mass segregation - the observed tendency for more massive protostars to be more centrally concentrated than low-mass protostars in the same cluster (e.g. Hillenbrand 1997, Hillenbrand & Hartmann 1998). This tendency is also evident in young stellar groups with only a few tens of members (Kirk & Myers 2011). If such mass segregation is primordial, it may be useful to compare the predictions of model luminosity distributions for inner and outer parts of young clusters, and for relatively younger and older clusters.

On the other hand, it may be possible to develop mass segregation dynamically on shorter time scales than previously believed (McMillan et al. 2007, Allison & Goodwin 2011). If so, the present model would not need to account for primordial mass segregation.

4.2. Isolated and clustered cluster models

Paper 1 described isolated condensation models which can represent initial conditions for protostars in the parts of clusters where effects of protostar crowding, dynamical processes, and global gravity are negligibly small. The condensations observed in filamentary lanes extending from some cluster centers appear sufficiently well-separated to be described by these isolated models. However in more crowded central regions, initial condensations for massive protostars could face a "spacing problem" if they have simple shape, and if their radii are comparable to the spacing of protostars in dense cluster centers, typically a few 0.01 pc. To make a star of mass 10 $M_\odot$ with efficiency 0.5 from a spherical initial condensation of diameter 0.01 pc would require an initial mean column density of ~50 g cm$^{-2}$, much greater than the value ~1 g cm$^{-2}$ inferred from observations of regions of massive star formation (Paper 1). Thus "clustered" cluster models as in this paper appear most useful for the central regions of young clusters, where massive protostars are forming, and where the typical spacing of protostars is a few 0.01 pc.

The degree of isolated and clustered star formation within a cluster may also depend on the relative importance of self-gravity and turbulent driving. With relatively strong turbulent driving, many simulations indicate more isolated star formation and fewer dynamical interactions. With weaker turbulent driving, or after initial turbulence has dissipated, the importance of global gravity, dynamical interactions, and competitive accretion increases (S. Offner, personal communication).

4.3. Do cores set protostar mass?

The models of this paper follow the idea that the IMF is set by processes which limit the time duration of gravitational accretion, rather than by initial conditions which limit the spatial extent of accretion. Such initial conditions include the sizes and masses of observed dense cores, or of condensations described by the Jeans mass or the Bonnor-Ebert mass. These observations and models describe many features of star-forming clouds, and it is generally accepted that protostars begin their accretion in dense cores. But it seems doubtful that dense cores can, by their structure alone, set the durations of accretion, or the distribution of protostar masses. This section discusses relations between accretion durations, dense cores, and the IMF.

4.3.1. *Jeans mass and the IMF.* Many authors have suggested that the characteristic stellar mass ~ 0.2 $M_\odot$ which defines the peak of the IMF arises from thermal fragmentation of dense star-forming gas (Larson 1985, Bonnell et al. 1997). This suggestion is supported by estimates of the temperature ~ 10 K and the density ~ $10^5$ cm$^{-3}$ typical of dense cores found in regions of isolated low-mass star formation (Jijina, Myers & Adams 1999, Enoch et al 2006). For these properties, the corresponding Jeans mass is of order 1 $M_\odot$ depending on the definition of Jeans mass used (McKee & Ostriker 2007). Taking into account the frequent association of protostars with dense cores (Beichman et al 1986) and a star formation efficiency of order 0.3 (Alves et al. 2007), it seems plausible that the modal mass of the IMF could arise from a characteristic core mass.

This similarity of the modal mass and the Jeans mass does not explain the physical basis of the star formation efficiency (SFE). Values of SFE which include 0.3 have been obtained by models of outflow feedback (Matzner & McKee 2000), Cunningham et al. 2011, Machida & Matsumoto 2011). However, these models do not predict an ensemble of cores having a mass distribution which resembles the IMF, shifted in mass by a factor equal to the SFE.

Furthermore, if the typical Jeans mass of star-forming gas accounted for the modal mass of the IMF, one might expect the range of such Jeans masses to account for the range of masses of the IMF, which exceeds a factor of $10^3$. Yet the temperatures and densities usually attributed to star-forming gas provide a range of Jeans masses much smaller than $10^3$. Furthermore, the thermal Jeans mass may be too simple a description of fragmentation even for low-mass cores, when the roles of rotation, turbulence, and magnetic fields are included realistically, and when the dynamical interaction of a core with its environment is taken into account.

4.3.2. CMF and IMF. The leading explanation of how core properties account for the IMF is the application of a constant star formation efficiency to observed distributions of core masses (CMF; Motte et al. 1998, Alves et al. 2007). But claims of a genetic relation between the CMF and the IMF have also been questioned because in crowded clusters, cores blend in projection (Kainulainen et al. 2009, Michel et al. 2011), because some cores can fragment and form multiple protostars (Hatchell & Fuller 2008), and because small cores may disperse before they make any stars (Myers 2009).

In this paper the IMF is determined by the rate and duration of accretion, and the mode of the IMF is set by the constant component of the mass accretion rate, and the typical accretion duration. This typical duration can be set by many factors, including ejection, competition, and dispersal due to stellar feedback, as discussed in Section 2.3.

4.3.3. Core boundaries. It seems unlikely that initial core structure can provide a useful limit on the duration of accretion. The typical observed core is centrally condensed, so its free-fall time increases with radius. If a core has a well-defined mass boundary, as in the model of a pressure-truncated isothermal sphere (Bonnor 1956, Ebert 1955), the free-fall time at the boundary sets the maximum duration of accretion. But this maximum duration applies only if the collapse is not limited sooner, by environmental and dynamical processes discussed above.

Furthermore, there is little evidence that most star-forming cores have boundaries which limit accretion. Some isolated cold dense globules are surrounded by hot, rarefied gas, such as B68 in the Loop I superbubble of the Sco-Cen OB association (Alves et al. 2001) or Thackeray 3 in the H II region IC2944 (Reipurth et al. 2003). Their images have visibly sharp boundaries at their phase transitions, and they appear to be well-described as cores with boundaries to mass flow.

In contrast, the environments of most star-forming cores do not resemble those of B68 and Thackeray 3. Star-forming cores in well-studied nearby complexes are surrounded by cold molecular gas, having lower density and greater velocity dispersion, and some filamentary structure. Such clump gas is evident in dust maps and images which do not filter out or subtract extended "background" structure, such as those obtained in studies of Lupus (Teixeira et al. 2005), or Perseus (Kirk et al. 2006). Clump gas dominates in many regions observed in spectral line emission by lines sensitive to gas density of a few $10^3$ cm$^{-3}$, such as the $J = 1$-0 transition of $^{13}$CO. Clump gas is the main mass reservoir in nearby cluster-forming regions, with typical mass $\sim 10^3$ $M_\odot$ (Ridge et al. 2003, Bergin & Tafalla 2007). The density profile of clump gas which surrounds cores appears shallower than that of the core gas, but does not indicate a core-clump boundary which could limit mass flow (Teixeira et al. 2005).

4.3.4. Isolated core interpretation. Despite the foregoing arguments, it is possible to assume that dynamical and environmental limits on accretion duration are negligible. Then the distribution of accretion durations assumed here can be interpreted as a distribution of infall times for an ensemble of isolated, bounded condensations, whose enclosed masses match the IMF with a constant star formation efficiency. If such condensations are spherically symmetric, their radial density profile is described by the two-component profile obtained in M10 and used in Paper 1. These model condensations therefore have a well-defined distribution of initial radii which can be compared with observations of cores.

The distribution of radii $\Phi_R = Rp(R)$, where $p(R) = p(m)dm/dR$, is obtained by combining expressions for the continuous Kroupa IMF (M10, equation (A1)), the enclosed condensation mass $M$ as a function of radius $R$ (Paper 1, equation (2)), and the definition of star formation efficiency $\varepsilon = m/M$, where $m$ is the protostar mass. Then

$$\Phi_R = \frac{aR(\alpha + \beta R^{4/3})}{\mu_{ps}^b (1 + \mu_{ps}^c)} \qquad (28)$$

where the normalized protostar mass is $\mu_{ps} = \varepsilon m/m_n$, and where the constants have values $a = 0.636$, $b=0.30$, $c=2.05$, $\alpha = 24.5$ $M_\odot$ pc$^{-1}$, $\beta = 834$ $M_\odot$ pc$^{-7/3}$, $\varepsilon = 0.30$, and $m_n = 0.205$ $M_\odot$.

This distribution in equation (28) was compared to the distribution of 97 starless core radii in Ophiuchus, the nearest cluster-forming region, based on 870 $\mu$m observations of the dust continuum with FWHM beam width 15 arcsec = 0.0091 pc at a distance of 125 pc (Sadavoy et al. 2010). This is the highest-resolution survey available which reports core radii for a significant sample in a cluster-forming region. These two distributions are compared in Figure 5.

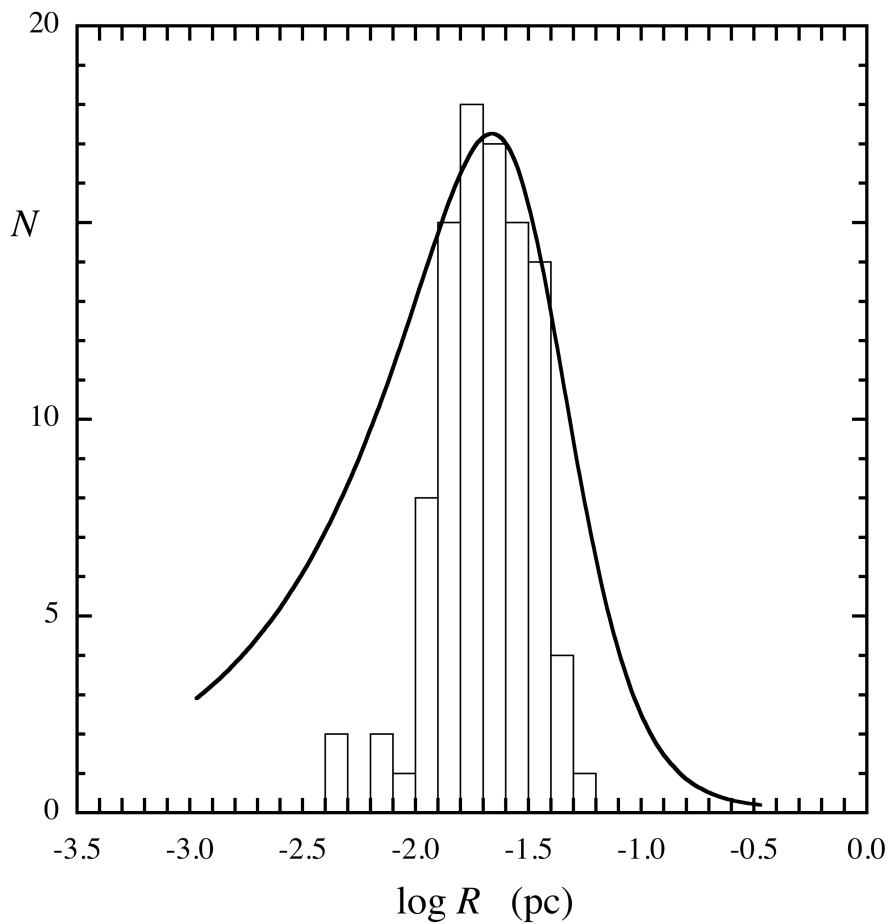

**Figure 5.** Observed and predicted distributions of core radii. *Histogram*, observed distribution of 97 core radii in a high-resolution survey of the nearest cluster-forming region, in Ophiuchus (Sadavoy et al. 2010). *Curve*, distribution of core radii required by equation (28), when the assumed distribution of accretion durations is interpreted as a distribution of core free-fall times and when the distribution of core masses matches the IMF with a star formation efficiency 0.3. The predicted distribution is significantly broader than the observed distribution.

These observed and model distributions of radii each have a single peak, and these peaks coincide at radius 0.020 pc for values of efficiency $\varepsilon$ near 0.3. However the model distribution is significantly broader than the observed distribution, by a factor 2.5 in its FWHM. The observed cores have a deficit of both small and large cores compared to the model distribution.

Some of the deficit of small observed cores can be ascribed to beam dilution, and to blending if the cores are projected sufficiently close to each other (Kainulainen et al. 2010, Michel et al. 2011). On the other hand some of this deficit is underestimated because some small starless cores may disperse before they can form stars.

It is more difficult to reconcile the deficit of large observed cores. The Oph sample has no observed cores with radius greater than 0.060 pc or mass greater than 5.2 $M_\odot$ (Sadavoy et al. 2010). With efficiency $\varepsilon = 0.3$ this implies a relatively low maximum protostar mass 1.6 $M_\odot$, corresponding to spectral type F0. Yet there are five stars with spectral types earlier than F0 already known in the Oph cluster, and two of these have spectral type B2, indicating stellar mass greater than 8 $M_\odot$ (Wilking et al. 2008). Evidently the Oph cluster has already made stars significantly more massive than it could form from its currently observed cores, if each core were required to form a single star with constant efficiency and no accretion of surrounding gas.

Thus in the Oph cluster, small and large cores do not appear in sufficient numbers to match the distribution expected from the present model of core-clump accretion with equally likely stopping, in the isolated core interpretation.

In contrast, this discrepancy between expected and observed core sizes does not occur when cores are embedded in an extended environment of clump gas which is available for accretion, and when the distribution of accretion durations is set by dynamical and environmental processes rather than by initial core boundaries. Then sufficient clump gas is available to make the most massive stars, and the distributions of core sizes and masses are not constrained to match those set by the IMF.

## 4.4. Massive stars are clump-fed

An important feature of the two-component accretion model is that the mass of low-mass stars is accreted primarily at the constant component of the accretion rate, while the mass of massive stars is accreted primarily at the mass-dependent component of the accretion rate, as discussed in Section 2.1. The implication of this property depends on the spatial model associated with the accretion rate. In the spatial model favored here, the central gas of a young cluster consists of primarily thermal cores embedded in a "clump" consisting of filaments and inter-filament gas, having lower density and more turbulence than the cores. The cores extend into their clump environment with no barrier to mass accretion.

In this picture it is plausible to associate the constant component of the accretion rate with infall of thermal core gas, as in the standard model of isolated low-mass star formation (Shu et al. 1987). With this association, and with equally likely stopping of accretion, such thermal core infall accounts for low-mass star formation and for the peak of the IMF (Myers 2009). However it cannot account for massive star formation and the high-mass tail of the IMF because it provides too few massive stars. Instead the mass-dependent component of the accretion rate is needed, and the mass which accretes at this rate must come from the clump gas which surrounds the thermal cores. In this picture, massive stars are clump-fed, as also indicated by the simulations of Smith et al. (2009) and of Wang et al. (2010).

An alternate spatial model is the isolated core picture discussed above in Section 4.3.4. There, each protostar arises from a well-defined initial core of sufficient mass and size to make that protostar with constant mass efficiency. The distribution of initial core masses matches the IMF, and the distribution of infall durations is equal to the distribution of core infall times. The clump gas does not contribute to star formation.

In this picture the initial core radius needed to produce the most massive protostars having 8 $M_\odot$ in the Oph cluster is 0.21 pc. This radius is similar to the radius of the outermost contours of the cluster-forming clumps A and E in the Oph cloud, which each enclose numerous cluster members (Wilking et al. 2008, Figure 3). Thus the isolated cores needed to produce massive stars are so large that they more closely resemble cluster-forming clumps than the thermal cores which form low-mass stars.

The isolated core model may therefore be useful in the outermost filamentary regions of young clusters, where protostars are formed with relatively low surface density, and where massive protostars are absent (Paper 1). However the present model of multiple cores in a clump which can provide accreting gas appears to have advantages for the denser parts of young clusters, which form both massive and low-mass stars with high density.

4.5. Cluster and protostar ages

When the mass exponent $p$ has its best-fit value 1.2-1.3 the accretion duration for a protostar of given mass is not constant as for $p=2$ but increases monotonically with time, more nearly as for $p = 1$ in Figure 1. This property implies that a massive protostar has had a greater accretion duration than a less massive protostar, i.e the massive protostar was "born" earlier. Consequently, a more luminous protostar was born before a less luminous protostar. This property allows an estimate of the age of the oldest protostar in a cluster, from equations (5) and (9).

The age of the oldest YSO in a cluster can also be estimated in this model, from the ratio of protostars to YSOs in equation (20). It may be useful to compare the accretion age and cluster age based on these estimates with cluster ages based on evolutionary tracks for pre-main sequence stars.

4.6. Protostar luminosity distribution

The luminosities of protostars have been of particular interest since observations from the *IRAS* satellite indicated that the typical protostar in Taurus has bolometric luminosity lower by a factor 10-50 than expected from the "standard" model of steady spherical protostar accretion from a singular isothermal sphere (Kenyon et al. 1990).

With more extensive and more sensitive observations from the *Spitzer Space Telescope*, this "luminosity problem" remains significant (Dunham et al. 2010, OM11). The discrepancy is often attributed to the idea that the accretion from the disk to the protostar is more episodic than the infall from the core to the disk, and that the fraction of an episodic cycle spent in accretion from the disk to the protostar is relatively small (Kenyon et al. 1990, Vorobyov & Basu 2005).

As more data have become available, it has become possible to compare distributions of observed and model luminosities, giving a test which is more detailed than comparison of typical values such as the mean, median, or mode. Section 3 compares the model luminosity distribution to a compilation of 112 luminosities in nearby star-forming clouds, based on integration over the broadband spectra of 39 Class 0 and 73 Class I protostars observed primarily in the Serpens, Ophiuchus, and Perseus complexes. The observations were obtained primarily with the *Spitzer Space Telescope* under the "c2d" program (Evans et al. 2009) and with the "bolocam" array at the Caltech Submillimeter Observatory (Enoch et al. 2009).

This observed luminosity distribution was compared to models of a collapsing singular isothermal sphere, following Young & Evans (2005), by Dunham et al. (2010). It was also compared to models of a collapsing isothermal sphere, to the turbulent core model of McKee & Tan (2003), and to an analytic version of the competitive accretion model of Bonnell et al. (1997) by OM11.

OM11 found that among the models they considered, the best fit to the data of Dunham et al (2010) was provided by what they term the two-component competitive accretion model (2CCA), followed closely by the two-component turbulent core model (2CTC), which is a blend of the isothermal sphere and turbulent core models. The 2CCA model is an analytic approximation to the competitive accretion simulations of Bonnell et al. (1997), Bonnell et al. (2001), and Smith et al. (2009). The dependences of mass accretion rate on protostar mass for the 2CTC and 2CCA models, given in equations (14) and (15) of OM11, are similar to that adopted independently in equations (1) and (2) of this paper, with $p = 1.2\text{-}1.3$.

The best-fit value of the luminosity accretion efficiency determined in this paper, $\gamma = 0.5$, is in excellent agreement with the corresponding value determined by OM11. OM11 attribute this factor, which they term $f_{acc,eff}$, to two aspects of disk accretion, an efficiency of 0.75 due to non-radiative energy loss in winds, and an efficency of 0.25 coresponding to a modest amount of episodic accretion. Together these yield an overall efficiency of $f_{acc,eff} = 0.56$.

It appears that two-component accretion models of the type considered here and by OM11 can account for the distribution of protostar luminosities in nearby star-forming regions, in a way which effectively resolves the luminosity problem described by Kenyon et al. (1990).

4.7. Relation to recent models

The present model is an analytic model of time-dependent accretion which specifies core and clump components of accretion rate, and which also specifies when accretion stops. The model predicts a time-dependent protostar mass function (PMF), which tends to a time-independent final mass function for times long compared to the mean stopping time. The parameters of the final protostar mass function are adjusted to match the IMF of stars. Similarly, the model predicts a time-dependent protostar accretion luminosity function (PLF), which tends to a time-independent form for late times. This function is matched to an observed distribution of protostar luminosities.

This model differs from recent accretion models which match the IMF but not the PLF, and it differs from models which assume the IMF as an input and then predict the PLF. It is believed that no published accretion model predicts and matches both the IMF and a PLF, as the present model does. Similarly, no published model of stochastic stopping of accretion predicts and matches both the IMF and a PLF.

Recent models which are most similar to the present model in their accretion rates and in their relation to the IMF are the 2CTC and 2CCA models of Offner & McKee (2011, OM11) and of McKee & Offner (2010; MO10), and the core-clump collapse models of Myers (2009; M09) and Myers (2011; Paper 1).

Nonetheless, the present model differs from those of OM11. This model predicts and matches a mass function to the IMF, while OM11 assume the IMF as an input. Both models model predict and compare PLFs to observations, but this model adjusts parameters for best match. The present PLF depends on time, while the OM11 PLFs are time-independent. This model specifies when accretion stops, due to a stochastic description of equally likely stopping. OM11 get a different distribution of durations for each accretion model they consider, from combining each accretion model with the IMF. The accretion rates in this model depend only on protostar mass, while those in OM11 depend on both mass and final mass. This model describes protostar masses up to 10 $M_\odot$, while OM11 consider masses up to 3 $M_\odot$.

The present model also differs from those of M09 and Paper 1 because its accretion is based on a two-component mass-dependent flow, appropriate to the complex structure in observations and simulations of dense young clusters. In contrast, the accretion in M09 and in Paper 1 assumes the isolated gravitational collapse of an initially static condensation of fixed spatial extent. The present results differ from those of M09 and Paper 1 because they predict distributions of both mass and luminosity, while M09 and Paper 1 predict only a mass distribution.

## 5. Conclusion

This paper presents an analytic model of time-dependent protostar accretion for application to the dense parts of young clusters, which accounts for observed distributions of protostar masses and luminosities. Its main features are:

1. The model is based on constant and mass-dependent components of the mass accretion rate.

2. Accretion stops stochastically, due to dynamical ejection, competition with other accretors, and dispersal of dense gas by stellar feedback.

3. Accretion flows of short duration resemble Shu accretion and make low-mass stars, while long flows resemble reduced Bondi accretion and make massive stars.

4. The model predicts a time-independent distribution of final protostar mass, and time-dependent distributions of protostar mass and accretion luminosity.

Two spatial models of initial structure are discussed. In each, protostars are born in cores, they have the same distribution of accretion durations, and their final masses match the IMF. In the core-clump model, multiple cores are embedded in an extended clump, whose gas is available for accretion onto the protostars. Accretion durations are limited by ejection, competition, and stellar feedback, and are described by equally likely stopping. In the isolated core model, accretion durations are due to the infalls of isolated cores, whose well-defined masses match the IMF with constant star formation efficiency.

It is shown that the isolated core model requires a distribution of initial core radii substantially broader than the distribution of 97 core radii in a high-resolution study of the nearest embedded cluster, in Ophiuchus. The model requires initial cores much larger and more massive than are observed, to account for massive stars following the IMF, or for massive stars already formed in the Oph cluster. Such large cores resemble cluster-forming clumps which harbor multiple stars, more than they resemble the cores which make individual low-mass stars.

This comparison favors the core-clump model over the isolated core model, for star formation in embedded clusters. It suggests that initial core structure need not set protostar mass, and that massive stars are clump-fed.

Two models of stellar birthrate are considered. If all the stars in a cluster are born at the same time ("coeval" star formation), the predicted distribution of accretion luminosity has a narrow spike at the highest luminosity, corresponding to all the protostars still accreting. This property does not match observed distributions of protostar luminosity.

If instead stars in a cluster have a constant birthrate, several useful conclusions follow from the model of core-clump accretion with equally likely stopping:

1. The ratio of YSOs to protostars increases approximately linearly with the star-forming age of an embedded cluster. Application to recent studies of YSO and protostar populations indicates star-forming ages of ~1 Myr in nearby embedded clusters having one protostar for every ten YSOs.

2. Protostar masses have a time-dependent distribution which tends toward the time-independent final mass distribution, as the cluster age exceeds the mean accretion duration.

3. Model parameters are adjusted so that the final protostar mass distribution matches the IMFs of Kroupa (2002) and Chabrier (2005). These parameters are the exponent of the clump mass accretion rate, $p = 1.2$; the core component of the mass accretion rate, $\dot{m}_{core} = 1.7 \times 10^{-6}\ M_\odot\ \mathrm{yr}^{-1}$; the time scale of the clump mass accretion rate, $\tau_{clump} = 0.2$ Myr; and the mean accretion duration, $\bar{t} = 0.1$ Myr.

4. Protostars have a time-dependent distribution of accretion luminosity, whose shape is similar to that of the mass distribution. This distribution also tends to a time-independent form, as cluster age exceeds the mean accretion duration.

5. The time-independent protostar luminosity distribution matches that in nearby star-forming regions (Dunham et al. 2010) provided the luminosity accretion efficiency is close to $\gamma = 0.5$. This result suggests that a modest amount of episodic accretion is sufficient to resolve the protostar luminosity problem, as found earlier by Dunham et al. (2010) and Offner & McKee (2011).

6. The parameters $\dot{m}_{core}$ and $\bar{t}$ set the peak of the IMF at $0.2\ M_\odot$, independent of the local Jeans mass.

7. The maximum protostar luminosity in a cluster indicates the age and mass of its oldest accreting protostar. Accretion luminosity of 6300 $L_\odot$ indicates an accretion age 0.49 Myr and a mass 10 $M_\odot$.

**Acknowledgements**

Helpful discussions are acknowledged with Fred Adams and Stella Offner, Shantanu Basu, Mike Dunham, Alyssa Goodman, Chris McKee, Ralph Pudritz, Howard Smith, Rowan Smith, Jonathan Tan, and Qizhou Zhang. The referee pointed out an inconsistency in the original version, whose correction led to a significant improvement. The referee also made helpful suggestions. Terry Marshall and Irwin Shapiro provided support and encouragement.